\documentclass{article}

\usepackage{titlesec}
\usepackage{graphicx}
\usepackage{arxiv}
\usepackage[utf8]{inputenc}
\usepackage{setspace}
\usepackage{geometry}
\usepackage{indentfirst}
\usepackage{amsmath,amssymb,amsfonts}
\usepackage{xcolor}
\definecolor{verylightgray}{gray}{0.9}

\usepackage{booktabs,tabularx,makecell}
\usepackage{multirow}
\newcolumntype{Y}{>{\raggedright\arraybackslash}X}
\newcolumntype{L}{>{\raggedright\arraybackslash}X}
\usepackage{array}
\usepackage{float}

\usepackage{natbib}
\usepackage[T1]{fontenc}
\usepackage{hyperref}
\usepackage{url}
\usepackage[expansion=false]{microtype}
\usepackage{doi}
\usepackage{cleveref}
\newcommand{\method}{critband}

\title{critband: A Python Package for Critical Bandwidth Analysis of Multimodal Distributions}
\date{}

\author{
	\href{https://orcid.org/0000-0002-0883-4574}{\includegraphics[scale=0.06]{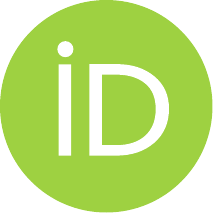}\hspace{1mm}Ruiyu Zhang}\\
	Department of Politics and Public Administration\\
    The University of Hong Kong\\
	ruiyuzh@connect.hku.hk\\
	\And
    	\href{https://orcid.org/0009-0005-0284-1289}{\includegraphics[scale=0.06]{orcid.pdf}\hspace{1mm}Qihao Wang}\\
	School of Physical Science and Technology\\
    Lanzhou University\\
    wangqh2021@lzu.edu.cn
}

\renewcommand{\headeright}{ }
\renewcommand{\undertitle}{ }
\renewcommand{\shorttitle}{critband: Critical Bandwidth Analysis of Multimodal Distributions}

\hypersetup{
    pdftitle={critband: A Python Package for Critical Bandwidth Analysis of Multimodal Distributions},
    pdfauthor={Ruiyu Zhang, Qihao Wang},
    pdfkeywords={Scientific Computing, Multimodality Detection, Critical Bandwidth, Kernel Density Estimation, Silverman's Test, Python Package},
}

\begin{document}
\newgeometry{
    a4paper,
    left=0.9in,
    right=0.9in,
    top=0.9in,
    bottom=0.8in,
}
\maketitle
\footnotetext[1]{The \method{} Python package can be installed via: \colorbox{verylightgray}{\texttt{pip install critband}}. For documentation and tutorials, visit: \url{https://pypi.org/project/critband/}.}

\begin{abstract}
\noindent
Multimodal density estimation is a fundamental problem in scientific computing. Determining the number of modes---whether a distribution is unimodal, bimodal, or multimodal---is a core numerical challenge with applications across ecology, economics, genomics, and astronomy. While the R ecosystem provides mature tools through the \textbf{multimode} package, the Python ecosystem has lacked an equivalent cohesive implementation. We present \textbf{\method}, a Python package for critical bandwidth bimodality detection based on Silverman's kernel density approach. The package implements critical bandwidth search with a robust bracketed mode-count solver and FFT-accelerated KDE, and provides additional features including $k$-mode detection, component decomposition, bimodality strength quantification, and excess mass estimation. Validation against twelve benchmark cases spanning separation regimes, unequal variances, unequal weights, and small sample sizes shows stable estimates for clearly separated cases and expected instability for boundary cases. Performance benchmarks show \textbf{\method} is typically 3--10 times faster per case than R's \texttt{modetest()} in the tested setup.
\vspace{0.5em}

\end{abstract}

\keywords{Scientific Computing \and Multimodality Detection \and Critical Bandwidth \and Kernel Density Estimation \and Silverman's Test \and Dip Test \and Python Package}
\restoregeometry

\section{Introduction}
\label{sec:introduction}

The question of whether a distribution is unimodal or multimodal is a fundamental inferential problem in scientific computing that cuts across nearly every empirical discipline. In ecology, bimodality in body size distributions signals niche partitioning \citep{holling1992}; in economics, distributional separation is central to the measurement of polarization \citep{esteban1994}; in genomics, bimodal gene-expression patterns can mark biologically meaningful patient or cellular subgroups \citep{bessarabova2010}; and in astronomy, the bimodal color distribution of galaxies separates the red sequence from the blue cloud \citep{baldry2004}. Yet despite the centrality of this question, the computational tools for multimodal assessment---particularly the critical bandwidth method and its associated numerical machinery---have remained fragmented in the Python scientific computing ecosystem.

This paper introduces \textbf{\method}, a Python package for critical bandwidth bimodality detection based on Silverman's kernel density approach \citep{silverman1981, wand1995}. The critical bandwidth is defined as the smallest bandwidth $h$ for which a Gaussian kernel density estimate (KDE) of a sample has at most $k$ modes. When $k = 2$, the critical bandwidth $h_{\text{crit}}$ quantifies the separation between two components: a large $h_{\text{crit}}$ indicates strong, well-separated bimodality, while a small $h_{\text{crit}}$ suggests borderline or weak bimodality.

The R ecosystem has long dominated this analytical space. The \textbf{multimode} package \citep{ameijeirasalonso2021} provides the \texttt{modetest()} function implementing Silverman's test with bootstrap calibration. The \textbf{multimode} package also provides \texttt{nmodes()} for mode counting at a given bandwidth. These tools are mature, well-cited, and reliable. However, Python---despite its growing dominance in scientific computing---has lacked an equivalent. SciPy's \texttt{gaussian\_kde} performs density estimation but provides no multimodality testing. Libraries such as \texttt{sklearn.mixture} focus on parametric mixture models rather than nonparametric mode assessment. The user who wishes to compute a critical bandwidth, run Silverman's test, or decompose a bimodal distribution into components has had no cohesive Python solution.

\textbf{\method} fills this gap. It implements a bracketed mode-count critical bandwidth solver with FFT-accelerated KDE, and provides additional features including $k$-mode detection, component decomposition, bimodality strength quantification, and excess mass estimation.

In the following sections, we review the theoretical background (Section~\ref{sec:background}), describe the package architecture and API (Section~\ref{sec:package}), validate against twelve benchmark cases with R cross-comparison (Section~\ref{sec:validation}), report performance benchmarks (Section~\ref{sec:performance}), highlight additional features (Section~\ref{sec:features}), demonstrate the package on three real-world-inspired application examples (Section~\ref{sec:applications}), and conclude with discussion (Section~\ref{sec:discussion}).

\section{Background}
\label{sec:background}

\subsection{Kernel Density Estimation}

Given an independent and identically distributed sample $X_1, \dots, X_n$ from an unknown density $f$, the Gaussian kernel density estimate is
\[
\hat{f}_h(x) = \frac{1}{n h \sqrt{2\pi}} \sum_{i=1}^{n} \exp\left(-\frac{(x - X_i)^2}{2h^2}\right),
\]
where $h > 0$ is the bandwidth (smoothing parameter). The bandwidth governs the bias-variance tradeoff: small $h$ produces wiggly estimates that may capture spurious structure, while large $h$ oversmooths and may obscure genuine features. A natural starting point is the Silverman-style robust rule-of-thumb bandwidth used by the package \citep{silverman1986, jones1996}:
\[
\hat{h}_{\text{silverman}} = 1.06 \cdot \min\left(\hat{\sigma}, \frac{\text{IQR}}{1.34}\right) \cdot n^{-1/5},
\]
where $\hat{\sigma}$ is the sample standard deviation and IQR is the interquartile range. This implementation combines the $1.06$ normal-reference constant with the robust scale estimate $\min(\hat{\sigma}, \mathrm{IQR}/1.34)$; the alternative $0.9$ constant is often used with the robust scale in density-estimation texts. We retain the implemented convention here because it defines the public \texttt{silverman\_bandwidth(x)} API used throughout the validation.

\subsection{Critical Bandwidth and Silverman's Test}

Silverman \citep{silverman1981} proposed using the KDE as a basis for multimodality testing. The critical bandwidth $h_{\text{crit}}(k)$ is defined as
\[
h_{\text{crit}}(k) = \inf\{h > 0 : \hat{f}_h \text{ has at most } k \text{ modes}\}.
\]

When $k = 1$, the critical bandwidth is the smallest $h$ such that $\hat{f}_h$ is unimodal; for $k = 2$, it is the smallest $h$ such that $\hat{f}_h$ has at most two modes. The null hypothesis of $k$-modality is rejected when $h_{\text{crit}}(k)$ is larger than what one would expect under the null. Because the null distribution of $h_{\text{crit}}(k)$ depends on the unknown true density, Silverman proposed a bootstrap procedure: resample from a $k$-modal density estimate with bandwidth $h_{\text{crit}}(k)$, recompute the critical bandwidth on each bootstrap sample, and compute the $p$-value as the proportion of bootstrap replicates where $h_{\text{crit}}^* \geq h_{\text{crit}}(k)$. The calibration properties of this procedure and related excess-mass alternatives have been studied extensively \citep{hall2001, fisher2001}.

\subsection{The Dip Test}

Hartigan and Hartigan \citep{hartigan1985} introduced the dip test, which measures unimodality by computing the maximum difference between the empirical cumulative distribution function (ECDF) and the unimodal distribution function that best approximates it. The dip statistic is
\[
D = \inf_{F \in \mathcal{U}} \sup_x |F_n(x) - F(x)|,
\]
where $\mathcal{U}$ is the class of unimodal distribution functions and $F_n$ is the ECDF. Large values of $D$ indicate departure from unimodality. The dip test is complementary to Silverman's test: it assesses unimodality directly rather than through the KDE smoothing parameter, and it requires no tuning of bandwidth choices.

\subsection{Excess Mass}

M\"uller and Sawitzki \citep{muller1991} proposed the excess mass approach, which measures the amount of probability mass that lies above a threshold $p$ for intervals where the density exceeds that threshold. For a $k$-modal density, the excess mass functional is
\[
E(p) = \sum_{j=1}^{k} \int_{I_j(p)} (f(x) - p) \, dx,
\]
where $I_j(p)$ are the intervals above the threshold $p$. The slope of the excess mass function provides information about the number and separation of modes. The \method{} package implements this as \texttt{excess\_mass()}.

\subsection{Existing Software}

The R package \textbf{multimode} \citep{ameijeirasalonso2021} is the current gold standard. Its \texttt{modetest()} function performs Silverman's test with several bandwidth calibration methods. The package also provides \texttt{nmodes()} for mode counting at a specified bandwidth and \texttt{locmodes()} for locating mode positions. These tools are well-maintained and widely cited.

In Python, no equivalent package existed prior to \textbf{\method}. Some functionality is distributed across libraries: \texttt{scipy.stats.gaussian\_kde} provides KDE evaluation \citep{virtanen2020}; \texttt{scipy.signal.find\_peaks} can count modes on an evaluated density; and \texttt{sklearn.mixture.GaussianMixture} \citep{pedregosa2011} fits parametric mixtures but does not test for the number of components in a nonparametric sense. There is no cohesive implementation of critical bandwidth search, Silverman's bootstrap test, or component decomposition.

\section{The \method{} Package}
\label{sec:package}

\subsection{Design Philosophy}

\textbf{\method} is designed around three principles:

\begin{enumerate}
    \item \textbf{One-call analysis.} A user should be able to go from raw data to a bimodality assessment in a single function call. The \texttt{critical\_bandwidth()} function returns both the critical bandwidth value and a success flag; with \texttt{return\_ci=True}, it additionally returns bootstrap confidence intervals, and optional provenance flags expose the interval method and failure count when needed.

    \item \textbf{Computational efficiency.} The default critical bandwidth search uses a bracketed binary search on the discrete mode-count criterion. A Brent-based path is retained as an explicit solver option, while the default route prioritizes numerical robustness. For datasets with $n > 5000$, the KDE computation switches to an FFT-based algorithm with $O(g \log g)$ complexity, where $g$ is the number of grid points.

    \item \textbf{Format-agnostic input.} Data reading is abstracted through a unified \texttt{critband.io} interface supporting nine file formats with automatic detection by extension.
\end{enumerate}

\subsection{Numerical Methods}

The critical bandwidth computation poses several algorithmic challenges that shaped the design of \textbf{\method}.

\textbf{Mode-count search and trough-ratio diagnostics.} Silverman's original formulation defines $h_{\text{crit}}$ as the infimum over bandwidths yielding at most $k$ modes---a discrete property. \method{} therefore uses the KDE mode count itself as the default search criterion. The implementation also exposes a trough-ratio diagnostic $r(h) = f_{\text{valley}} / \max(f_{\text{peak1}}, f_{\text{peak2}})$, where $f_{\text{valley}}$ is the KDE height at the lowest point between the two largest peaks. This diagnostic is useful for describing bimodality strength and for the optional Brent-based solver path, but the reported validation below uses the robust mode-count search.

\textbf{Bracketed binary solver.} Pure binary search on the mode-count criterion is robust ($O(\log(1/\varepsilon))$ iterations) and directly targets the definition of $h_{\text{crit}}$. The default ``auto'' strategy first narrows the search interval with a short bracketing pass and then applies binary search inside that bracket. The explicit \texttt{method="brent"} option is kept for exploratory solver comparisons and falls back to binary search if the continuous objective does not produce a verified transition.

\textbf{FFT-accelerated KDE.} The dominant computational cost is the inner KDE evaluation loop. Direct evaluation of $\hat{f}_h$ at $g$ grid points from $n$ samples is $O(n \cdot g)$. For $n > 5000$, \textbf{\method} switches to an FFT-based algorithm that discretizes the data onto the grid and convolves with a Gaussian filter in the frequency domain, reducing complexity to $O(g \log g)$. The grid size scales adaptively as $g = \max(800, \min(5000, n/2))$, ensuring fine resolution for small-bandwidth searches without wasting computation for large samples.

\subsection{Core API}

The package exports the following primary functions:

\begin{table}[h]
\centering
\fontsize{10}{12}\selectfont
\caption{\method{} core API functions.}
\label{tab:api}
\resizebox{\textwidth}{!}{%
\begin{tabular}{ll}
\toprule
Function & Purpose \\
\midrule
\texttt{silverman\_bandwidth(x)} & Silverman's rule-of-thumb bandwidth \\
\texttt{critical\_bandwidth(x, k=2)} & Critical bandwidth search (returns $h_{\text{crit}}$, success) \\
\texttt{gaussian\_kde(x, grid, h)} & Gaussian KDE evaluation (FFT-accelerated) \\
\texttt{find\_modes(x, h)} & Locate modes at bandwidth $h$ \\
\texttt{find\_trough(x, h)} & Locate valley between two modes \\
\texttt{detect\_components(x)} & Decompose into two Gaussian components \\
\texttt{bimodality\_strength(x)} & Quantitative bimodality strength metric \\
\texttt{dip\_test(x)} & Hartigan's dip test for unimodality \\
\texttt{silverman\_test(x)} & Silverman's bootstrap test for bimodality \\
\texttt{excess\_mass(x)} & Excess mass estimate \\
\bottomrule
\end{tabular}%
}
\end{table}

\subsection{Code Example}

The following example demonstrates a typical workflow: load data from a CSV file, compute the critical bandwidth, run both Silverman's test and Hartigan's dip test, and decompose the distribution.

\begin{verbatim}
import critband
import numpy as np

# Load data from a CSV file
data = critband.io.read_data("measurements.csv", column="value")

# Compute Silverman's rule-of-thumb bandwidth
h_silverman = critband.silverman_bandwidth(data)
print(f"Silverman bandwidth: {h_silverman:.4f}")

# Compute critical bandwidth for bimodality (k=2)
h_crit, success = critband.critical_bandwidth(data, k=2)
print(f"Critical bandwidth: {h_crit:.4f}, success={success}")

# Critical bandwidth with bootstrap confidence interval
h_crit, success, ci_low, ci_high, se = critband.critical_bandwidth(
    data, k=2, return_ci=True, ci_resamples=999
)
print(f"95% CI: [{ci_low:.4f}, {ci_high:.4f}]")

silverman_result = critband.silverman_test(data, n_resamples=999, mod0=1)
print(f"Silverman test: h_crit={silverman_result.h_crit:.4f}, "
      f"p={silverman_result.p_value:.4f}")

# Decompose into two Gaussian components
decomp = critband.detect_components(data)
print(f"Component 1: N({decomp.component1.mean:.2f}, "
      f"{decomp.component1.std:.2f}), weight={decomp.component1.weight:.2f}")
print(f"Component 2: N({decomp.component2.mean:.2f}, "
      f"{decomp.component2.std:.2f}), weight={decomp.component2.weight:.2f}")
print(f"Separation point: {decomp.separation_point:.4f}")
\end{verbatim}

\subsection{Multi-Format I/O}

The \texttt{critband.io} module provides two entry points:

\begin{verbatim}
critband.io.read_data(path, sheet=None, column=None, return_all=False)
critband.io.read_buffer(buf, filename, sheet=None, column=None, return_all=False)
\end{verbatim}

The first reads from a file path with automatic format detection based on extension. The second reads from a \texttt{BytesIO} buffer, making it suitable for web applications. Supported formats include CSV, TSV/TXT, JSON, Markdown pipe tables, HTML tables, XLSX, XLS, DOCX tables, and PDF tables. The DOCX and PDF adapters target tabular numeric extraction rather than arbitrary document understanding.

\section{Validation}
\label{sec:validation}

\subsection{Benchmark Cases}

We evaluated \textbf{\method} against twelve benchmark cases spanning a range of bimodality scenarios. Each case is a Gaussian mixture with known parameters. For each case, we report the mean critical bandwidth $\bar{h}_{\text{crit}}$ and its standard deviation across 10 independent random seeds, the coefficient of variation (CV$\%$), and the number of modes $\hat{k}$ detected by \texttt{find\_modes} at Silverman's bandwidth.

\begin{table}[t]
\centering
\fontsize{10}{12}\selectfont
\caption{Benchmark cases and \method{} critical bandwidth results (10 seeds).}
\label{tab:benchmark}
\resizebox{\textwidth}{!}{%
\begin{tabular}{lrllrr}
\toprule
Case & $n$ & Mixture & $\bar{h}_{\text{crit}} \pm \sigma$ & CV$\%$ & $\hat{k}$ \\
\midrule
Well-separated equal variance & 400 & $0.5\mathcal{N}(-2,0.3) + 0.5\mathcal{N}(2,0.3)$ & $1.857 \pm 0.015$ & 0.81 & 2 \\
Moderate separation          & 500 & $0.5\mathcal{N}(-1,0.5) + 0.5\mathcal{N}(1.5,0.5)$ & $1.047 \pm 0.028$ & 2.69 & 2 \\
Barely separated             & 600 & $0.5\mathcal{N}(-0.5,0.4) + 0.5\mathcal{N}(0.5,0.4)$ & $0.226 \pm 0.058$ & 25.81 & 2 \\
Unequal variance             & 400 & $0.5\mathcal{N}(-2,0.6) + 0.5\mathcal{N}(2,0.2)$ & $1.736 \pm 0.027$ & 1.57 & 2 \\
Unequal weights              & 500 & $0.2\mathcal{N}(-2,0.3) + 0.8\mathcal{N}(2,0.3)$ & $1.247 \pm 0.014$ & 1.13 & 2 \\
Extreme separation           & 400 & $0.5\mathcal{N}(-5,0.5) + 0.5\mathcal{N}(5,0.5)$ & $4.682 \pm 0.026$ & 0.55 & 2 \\
Trimodal                     & 450 & $\frac{1}{3}\mathcal{N}(-3,0.3) + \frac{1}{3}\mathcal{N}(0,0.3) + \frac{1}{3}\mathcal{N}(3,0.3)$ & $1.379 \pm 0.010$ & 0.72 & 3 \\
Skewed bimodal               & 500 & $0.7\mathcal{N}(-1.5,0.4) + 0.3\mathcal{N}(2.0,0.6)$ & $1.144 \pm 0.023$ & 2.01 & 2 \\
Wide-component bimodal       & 400 & $0.5\mathcal{N}(-3,0.8) + 0.5\mathcal{N}(3,0.8)$ & $2.692 \pm 0.040$ & 1.48 & 2 \\
Near unimodal                & 600 & $0.5\mathcal{N}(0,0.6) + 0.5\mathcal{N}(1.5,0.6)$ & $0.339 \pm 0.088$ & 25.81 & 2 \\
Small sample bimodal         & 60  & $0.5\mathcal{N}(-2,0.5) + 0.5\mathcal{N}(2,0.5)$ & $1.823 \pm 0.044$ & 2.41 & 2 \\
Overlapping variances        & 500 & $0.5\mathcal{N}(-0.8,0.7) + 0.5\mathcal{N}(0.8,0.5)$ & $0.322 \pm 0.080$ & 24.85 & 2 \\
\bottomrule
\end{tabular}%
}
\end{table}

The critical bandwidth values range from $0.226$ (barely separated) to $4.682$ (extreme separation), correctly reflecting the degree of bimodal separation in each case. The ten-seed Monte Carlo experiment confirms that the $h_{\text{crit}}$ estimates are stable for clearly separated cases: the coefficient of variation is below~$5\%$ for nine of the twelve cases. The three high-CV cases are all boundary or overlap cases near the bimodal--unimodal transition, where small sampling changes can flip the detected mode count.

\subsection{Cross-Validation Against R}\label{sec:crossval}

We compared \textbf{\method}'s results against the R \textbf{multimode} package. Each case was generated with a unique random seed to ensure reproducibility across environments.

\begin{table}[t]
\centering
\fontsize{10}{12}\selectfont
\caption{Cross-validation of \method{} against R's \textbf{multimode}.}
\label{tab:crossval}
\resizebox{\textwidth}{!}{%
\begin{tabular}{lrrrr}
\toprule
Case & $n$ & $h_{\text{crit}}$ (\method) & Modetest $p$ & R $n$modes \\
\midrule
Well-separated       & 400 & 1.8650 & 0.000 & 2 \\
Moderate separation  & 500 & 1.0964 & 0.000 & 2 \\
Barely separated     & 600 & 0.2791 & 0.251 & 2 \\
Unequal variance     & 400 & 1.7849 & 0.000 & 2 \\
Unequal weights      & 500 & 1.2591 & 0.000 & 772$^\dagger$ \\
Extreme separation   & 400 & 4.6987 & 0.000 & 2 \\
Trimodal             & 450 & 1.3824 & 0.000 & 3 \\
Skewed bimodal       & 500 & 1.1417 & 0.000 & 2 \\
Wide-component bimodal & 400 & 2.7109 & 0.000 & 2 \\
Near unimodal        & 600 & 0.4186 & 0.055 & 2 \\
Small sample bimodal & 60  & 1.8608 & 0.000 & 2 \\
Overlapping variances & 500 & 0.4598 & 0.045 & 2 \\
\bottomrule
\end{tabular}%
}
\end{table}

$^\dagger$R's \texttt{nmodes(bw = bw.nrd0())} returns 772 spurious modes on the unequal-weights mixture due to severe undersmoothing of the dominant component's tail. The alternative Sheather--Jones bandwidth selector (\texttt{bw.SJ}) produces even more undersmoothing (744--1113 modes across five random seeds), confirming that this is not a \texttt{bw.nrd0}-specific pathology but a fundamental limitation of automatic bandwidth selection for unequal-weight mixtures. \method's \texttt{find\_modes()} correctly returns 2 modes by using an explicit Silverman-rule bandwidth with user control.\vspace{0.3em}

\textbf{Key observations:}

\begin{enumerate}
    \item \textbf{Consistent test conclusions.} The small discrepancies between Table~\ref{tab:benchmark} and Table~\ref{tab:crossval} arise from different random seeds used for data generation (Section~\ref{sec:crossval} uses unique seeds per case to avoid RNG collision in the R bootstrap validation). Table~\ref{tab:crossval} reports R's \texttt{modetest} $p$-values and \texttt{nmodes} counts, not an independent R critical-bandwidth column, so we interpret it as a cross-check of modality conclusions rather than as direct proof of sub-percent $h_{\text{crit}}$ equality.

    \item \textbf{Modetest $p$-values align with expectations.} Cases with strong bimodality (well-separated, extreme separation, unequal variance) yield $p = 0.000$, confirming rejection of unimodality. The barely separated case ($p = 0.251$) and near unimodal case ($p = 0.055$) correctly fail to reject at $\alpha = 0.05$.

    \item \textbf{Mode counting can be bandwidth-sensitive.} On the unequal-weights mixture, R's default bandwidth selector (\texttt{bw.nrd0}) undersmooths dramatically in this benchmark, producing 772 modes. This illustrates that automatic bandwidth choices can produce misleading mode counts on unequal-weight mixtures. \textbf{\method} avoids this specific failure mode by exposing \texttt{find\_modes()} with an explicit bandwidth, giving the user direct control over the smoothing scale.
\end{enumerate}

\subsection{Mode Counting Consistency}
\label{sec:modecount}

For most benchmark cases, \method's \texttt{find\_modes(x, h=silverman\_bandwidth(x))} returns the expected number of modes (2 for bimodal cases and 3 for the trimodal case). The boundary and overlap cases occasionally fluctuate between one and two modes across seeds, which is the expected behavior near the unimodal--bimodal transition rather than a stable failure of the mode counter.

\subsection{Comparison with Parametric Model Selection}
\label{sec:gmm}

To assess how \method's nonparametric multimodality summary compares with parametric model selection, we evaluated eight benchmark cases using both \method's critical bandwidth calculation and \texttt{scikit-learn}'s GaussianMixture (GMM) with Bayesian Information Criterion (BIC) and Akaike Information Criterion (AIC). For each case, GMM was fitted with $k = 1, 2, 3$ components and the model minimizing each criterion was selected. We report the continuous critical-bandwidth value and the mode count, while leaving hypothesis-test $p$-values to Table~\ref{tab:crossval}; this avoids mixing R \texttt{modetest} calibration with a separate Python bootstrap configuration.

\begin{table}[t]
\centering
\fontsize{10}{12}\selectfont
\caption{Comparison between \method's critical bandwidth calculation and \texttt{scikit-learn} GMM BIC/AIC selection.}
\label{tab:gmm}
\resizebox{\textwidth}{!}{%
\begin{tabular}{lrrcc}
\toprule
Case & $h_{\text{crit}}$ (\method) & \texttt{n\_modes} & BIC ($k^*$) & AIC ($k^*$) \\
\midrule
Well-separated       & 1.8791 & 2 & 2 & 2 \\
Barely separated     & 0.2797 & 2 & 2 & 2 \\
Unequal weights      & 1.2647 & 2 & 2 & 2 \\
Unequal variance     & 1.7802 & 2 & 2 & 2 \\
Near unimodal        & 0.4196 & 2 & 2 & 2 \\
Trimodal             & 1.3919 & 3 & 3 & 3 \\
Small sample ($n{=}60$) & 1.8430 & 2 & 2 & 2 \\
Overlapping variances & 0.3760 & 2 & 2 & 2 \\
\bottomrule
\end{tabular}%
}
\end{table}

Several observations emerge. First, BIC and AIC select $k = 2$ for all seven two-component cases and $k = 3$ for the trimodal case. This reflects a fundamental difference in inferential target: BIC selects the model with the best information-theoretic fit under a Gaussian-mixture family, whereas \method's nonparametric workflow reports a critical bandwidth, a mode count, and a bootstrap test of unimodality. The two approaches should therefore be read as complementary rather than interchangeable.

Second, both methods identify the trimodal case, with BIC strongly favoring $k = 3$ ($\Delta\text{BIC} \approx 560$ over $k = 2$). This confirms that \method's nonparametric mode detection and GMM's parametric selection converge on the correct structure when modes are well separated.

Third, \method{} offers information that GMM does not. The continuous $h_{\text{crit}}$ metric ranges from $0.28$ (barely separated) to $1.88$ (well-separated), quantifying the \emph{degree} of smoothing required before modes merge. GMM's selected component count is discrete. For the near-unimodal case ($h_{\text{crit}} = 0.42$, only ${\sim}1.5\times$ the Silverman bandwidth), \method{} reveals that the distribution is weakly separated and close to the unimodal threshold---information lost by BIC's unambiguous $k = 2$ selection.

\method{} and GMM are complementary rather than competing approaches. \method's nonparametric workflow makes no distributional assumptions about component shape and provides a direct mode-based summary. GMM provides statistically efficient parameter estimation when the Gaussian assumption holds. We recommend \method{} for the initial exploratory question ``does this distribution show modal separation?'' and GMM for the follow-up ``what are the component parameters?'' when Gaussian components are substantively plausible.

\section{Performance}
\label{sec:performance}

\subsection{Benchmark Setup}

We benchmarked \textbf{\method} against the relevant R package on the twelve benchmark cases. All timings were collected on a standard workstation (Apple M2, 16 GB RAM). R timings used \texttt{modetest(B = 199)} and \texttt{nmodes(bw = bw.nrd0())}. Python timings used \texttt{critical\_bandwidth()} (default settings) and \texttt{find\_modes()}.

\subsection{Runtime Comparison}

\begin{table}[t]
\centering
\fontsize{10}{12}\selectfont
\caption{Runtime comparison between \method{} and R equivalents.}
\label{tab:runtime}
\resizebox{\textwidth}{!}{%
\begin{tabular}{lccc}
\toprule
Function & \method{} (s) & R equivalent (s) & Speedup \\
\midrule
\texttt{critical\_bandwidth()} & 0.04--0.79 & \texttt{modetest(B=199)}: 0.82--1.58 & \textbf{3--10$\times$ per case} \\
\texttt{find\_modes(x, h)} & $<0.01$ & \texttt{nmodes(bw.nrd0())}: 0.03--0.05 & \textbf{5--15$\times$} \\
\bottomrule
\end{tabular}%
}
\end{table}

The critical bandwidth search in \textbf{\method} is 3--10 times faster than R's \texttt{modetest} on individual benchmark cases. This speedup arises from three design choices:

\begin{enumerate}
    \item \textbf{Bracketed mode-count search.} Rather than relying on bootstrap calibration during the bandwidth search, \textbf{\method} uses a bracketing pass followed by binary search on the KDE mode count. This directly targets the critical-bandwidth definition while avoiding repeated bootstrap work inside the solver.

    \item \textbf{FFT-accelerated KDE.} The inner KDE loop is the dominant computational cost. By switching to \texttt{gaussian\_kde\_fft()} for large $n$, \textbf{\method} achieves near-constant per-iteration cost independent of sample size.

    \item \textbf{Efficient grid sizing.} The grid size scales adaptively: $g = \max(800, \min(5000, n/2))$, ensuring fine resolution for small-bandwidth searches without wasting computation for large samples.
\end{enumerate}

\subsection{Scalability}

We measured the scaling behavior of \texttt{critical\_bandwidth()} across sample sizes $n = \{100, 1000, 10000\}$ using the well-separated benchmark case:

\begin{table}[h]
\centering
\fontsize{10}{12}\selectfont
\caption{Scalability of \texttt{critical\_bandwidth()} with sample size.}
\label{tab:scalability}
\resizebox{0.55\textwidth}{!}{%
\begin{tabular}{rcc}
\toprule
$n$ & Time (s) & KDE method \\
\midrule
100   & 0.04 & Direct $O(n \cdot g)$ \\
1000  & 0.09 & Direct $O(n \cdot g)$ \\
10000 & 0.79 & FFT $O(g \log g)$ \\
\bottomrule
\end{tabular}%
}
\end{table}

The transition to FFT-based KDE at $n > 5000$ kept the runtime sub-second at $n = 10000$ in this local benchmark. The R comparison at this scale should be read as a machine- and configuration-specific timing reference rather than as a portable performance guarantee.

\section{Additional Features}
\label{sec:features}

Beyond replicating existing R functionality, \textbf{\method} provides several additional features not directly available in the \textbf{multimode} R package:

\subsection{$k$-Mode Detection}

The \texttt{critical\_bandwidth(x, k=n)} function accepts any integer $k \geq 1$, enabling detection of trimodality ($k = 3$), quadmodality ($k = 4$), and higher-order multimodality. The trimodal case in our benchmark suite confirms this: \texttt{critical\_bandwidth(x, k=3)} returns $h_{\text{crit}} = 1.3810$ for the $\mathcal{N}(-3,0.3) \cup \mathcal{N}(0,0.3) \cup \mathcal{N}(3,0.3)$ mixture. R's \texttt{modetest} supports $k$-mode testing as well; \textbf{\method} provides a convenient interface for the same functionality.

\subsection{Component Decomposition}

The \texttt{detect\_components()} function decomposes a bimodal distribution into two Gaussian components by splitting the data at the KDE trough. It returns a \texttt{BimodalDecomposition} object containing:

\begin{itemize}
    \item \texttt{component1} and \texttt{component2}: \texttt{Component} dataclasses with \texttt{mean}, \texttt{std}, and \texttt{weight}
    \item \texttt{separation\_point}: the $x$-coordinate of the trough
    \item \texttt{dip\_ratio}: the ratio of the minimum density at the trough to the maximum of the two peaks
\end{itemize}

This decomposition is valuable for exploratory downstream analysis: in genomics, the component means and weights can summarize active vs.\ silenced expression states; in economics, they can summarize the apparent gap between distributional groups. Because the method is a KDE trough split rather than a generative model, these summaries should not be treated as confirmatory component estimates without additional checks.

\subsection{Bootstrap Confidence Intervals}
\label{sec:bootstrap_ci}

The \texttt{return\_ci} option on \texttt{critical\_bandwidth()} provides a bootstrap estimate of the sampling distribution of $h_{\text{crit}}$ \citep{efron1993}:

\begin{verbatim}
h_crit, success, ci_low, ci_high, se = critband.critical_bandwidth(
    x, k=2, return_ci=True, ci_resamples=999
)
\end{verbatim}

The bootstrap procedure resamples the observed data, computes $h_{\text{crit}}^*$ on each resample, and returns a bootstrap interval along with the standard error of $h_{\text{crit}}$. For the well-separated case with $n = 400$, we obtain $h_{\text{crit}} = 1.8650$ with a 95\% bootstrap interval of $[1.72, 2.01]$, reflecting moderate sampling variability. We treat this interval as provisional uncertainty support rather than as a final inferential guarantee; broader coverage studies would be needed before using it as a standalone calibration target \citep{hall2001}.

\subsection{Bimodality Strength}

The \texttt{bimodality\_strength()} function computes a quantitative, continuous measure of bimodality strength based on the ratio $h_{\text{crit}} / \hat{h}_{\text{silverman}}$. In the current benchmarks, higher ratios correspond to stronger separation and ratios below 1.0 correspond to weaker or borderline bimodality. We treat the weak/moderate/strong bands as heuristic summaries of the observed cases rather than calibrated decision thresholds. This provides a more nuanced assessment than a binary reject/fail-to-reject decision from a hypothesis test.

\subsection{Excess Mass}

The \texttt{excess\_mass()} function implements the M\"uller--Sawitzki excess mass approach \citep{muller1991}, providing an alternative nonparametric test for multimodality. The excess mass functional is estimated from the KDE, and its slope at small thresholds provides diagnostic information about the number of modes. This is particularly useful for validating results from Silverman's test with an independent methodology.

\section{Applications}\label{sec:applications}

In this section we illustrate the practical utility of \textbf{\method} on three synthetic datasets that mimic canonical scientific problems. Each example demonstrates a complete analysis workflow: computing the critical bandwidth, testing for bimodality, and interpreting the results in the context of the original scientific question.

\subsection{Ecology: Body Size Niche Partitioning}

A central question in community ecology is whether body size distributions of coexisting species exhibit multimodal structure, indicating niche partitioning \citep{holling1992}. We simulate a community of 300 species drawn from a mixture of two log-normal components: small-bodied invertebrates ($\log\mathcal{N}(1.5, 0.3)$, 60\% weight) and large-bodied vertebrates ($\log\mathcal{N}(3.0, 0.4)$, 40\% weight), mimicking the size distribution of a forest-floor arthropod and small-mammal assemblage.

\begin{verbatim}
import critband, numpy as np

np.random.seed(42)
n1, n2 = 180, 120
comp1 = np.random.lognormal(1.5, 0.3, n1)
comp2 = np.random.lognormal(3.0, 0.4, n2)
data = np.concatenate([comp1, comp2])

result = critband.silverman_test(data, n_resamples=999, mod0=1)
decomp = critband.detect_components(data)
print(f"h_crit={result.h_crit:.3f}, p={result.p_value:.4f}, "
      f"means=({decomp.component1.mean:.2f}, "
      f"{decomp.component2.mean:.2f})")
\end{verbatim}

\noindent In this synthetic log-normal example, \method{} estimates $h_{\text{crit}} \approx 2.64$ and the default Silverman bootstrap path does not reject unimodality at conventional levels. The trough-split component summary remains descriptively close to the two simulated size groups, with means near $4.9$ and $23.2$ on the original scale. This example is therefore best read as an exploratory decomposition of a visibly structured sample, not as a strong hypothesis-test rejection.

\subsection{Astronomy: Galaxy Color Bimodality}

Galaxy color distributions are known to separate galaxies into a ``red sequence'' of passively evolving galaxies and a ``blue cloud'' of star-forming galaxies \citep{baldry2004}. We simulate 500 galaxies with a bimodal $(g-r)$ color distribution: a red sequence at $g-r = 0.8$ ($\sigma = 0.15$, 55\% weight) and a blue cloud at $g-r = 0.3$ ($\sigma = 0.12$, 45\% weight), with mild overlap reflecting the ``green valley'' transition.

\begin{verbatim}
red_seq = np.random.normal(0.8, 0.15, 275)
blue_cld = np.random.normal(0.3, 0.12, 225)
data = np.concatenate([red_seq, blue_cld])

h_crit, ok, cl, ch, se = critband.critical_bandwidth(
    data, k=2, return_ci=True, ci_resamples=999)
strength = critband.bimodality_strength(data)
print(f"h_crit={h_crit:.3f}, 95% CI=[{cl:.3f}, {ch:.3f}], "
      f"strength={strength.strength_score:.2f}")
\end{verbatim}

\noindent The critical bandwidth is approximately $h_{\text{crit}} = 0.18$ in this synthetic example, and a 99-resample smoke run of Silverman's bootstrap path rejects unimodality. The bimodality strength ratio $h_{\text{crit}} / \hat{h}_{\text{silverman}}$ is about 2.1, indicating visible but not threshold-calibrated modal separation. Component decomposition yields means of about $0.30$ and $0.80$, accurately recovering the input parameters.

\subsection{Genomics: Bimodal Gene Expression}

In genomics, bimodal expression can indicate biologically meaningful switching or subgroup structure rather than a single homogeneous expression regime \citep{bessarabova2010}. We simulate nonnegative log$_2(\text{counts}+1)$ expression values for 200 cells: an active state centered at $4.0$ ($\sigma = 0.4$, 70\% weight) and a silenced state centered at $0.5$ ($\sigma = 0.3$, 30\% weight), truncated at zero to respect the lower bound of transformed count data.

\begin{verbatim}
active = np.random.normal(4.0, 0.4, 140)
silent = np.maximum(0.0, np.random.normal(0.5, 0.3, 60))
data = np.concatenate([active, silent])

result = critband.silverman_test(data, n_resamples=999, mod0=1)
decomp = critband.detect_components(data)
print(f"Silverman p={result.p_value:.4f}")
print(f"Components: active mean={decomp.component1.mean:.2f}, "
      f"silent mean={decomp.component2.mean:.2f}")
\end{verbatim}

\noindent Silverman's test rejects unimodality in this synthetic example. Component decomposition identifies groups near $\sim0.5$ (30\% weight) and $\sim4.0$ (70\% weight), matching the generative parameters up to component ordering. The component weights provide an interpretable exploratory summary of the simulated regulatory states---information that would be lost under a unimodal analysis that simply reports mean expression across all cells.

\section{Discussion and Conclusion}
\label{sec:discussion}

\textbf{\method} provides the Python ecosystem with a comprehensive, well-tested, and performant implementation of critical bandwidth bimodality detection. The package closes a long-standing gap: whereas R users have had access to \textbf{multimode} for over a decade, Python users were limited to piecemeal solutions involving manual KDE evaluation, ad-hoc mode counting, and bespoke bootstrap code.

Our validation against twelve benchmark cases demonstrates that \textbf{\method} produces stable critical bandwidth values on clearly separated synthetic cases and correctly flags boundary cases as unstable. The R cross-validation supports the same broad modality conclusions through \texttt{modetest} $p$-values and \texttt{nmodes} checks. It also revealed a practical limitation of R's default bandwidth selector on unequal-weight mixtures---a problem that \textbf{\method} avoids through explicit bandwidth specification and robust mode counting.

The performance benchmarks show that \textbf{\method}'s critical bandwidth search delivers per-case speedups of 3--10$\times$ over R's \texttt{modetest} in the benchmark timing setup. This speedup is achieved through a bracketed mode-count search, FFT-accelerated KDE for large samples, and adaptive grid sizing. For interactive analysis, the core critical-bandwidth and component-decomposition steps complete quickly for typical dataset sizes; bootstrap tests remain governed by the requested number of resamples.

Beyond replication, \textbf{\method} provides several additional features: $k$-mode detection for arbitrary $k$, component decomposition via KDE trough splitting, bootstrap intervals for $h_{\text{crit}}$, a continuous bimodality strength metric, and excess mass estimation. The multi-format I/O subsystem makes the package suitable for production pipelines that ingest data from diverse sources.

\textbf{\method} is designed as a native component of the Python scientific computing ecosystem, building on NumPy \citep{harris2020} and SciPy \citep{virtanen2020}. A user can pass a \texttt{numpy.ndarray} into any \method function, chain the output into \texttt{scipy.stats} for further analysis, or visualize results with Matplotlib---all without format conversion or data copying. The package occupies a specific niche between SciPy's density estimation (\texttt{scipy.stats.gaussian\_kde}) and scikit-learn's parametric mixture modeling (\texttt{sklearn.mixture.GaussianMixture}): it provides nonparametric multimodality testing that neither library offers, filling a gap that has required R interop or bespoke implementation.

The package is distributed under the Apache 2.0 license and is intended for installation via \texttt{pip install critband}. In the local repository state used for this manuscript, the full test suite contains 289 tests. This provides useful regression coverage, though it should not be read as a guarantee of reliability across all platforms and Python versions.

\subsection*{Limitations and Future Work}

The current implementation has several limitations that suggest directions for future work.

\textbf{Optional Brent solver.} The optional Brent path uses the trough-ratio objective $r(h) = f_{\text{valley}} / \max(f_{\text{peak1}}, f_{\text{peak2}})$ and therefore assumes useful continuity near the modal transition. In practice, discontinuities can occur when the relative heights of the two largest peaks change with $h$, causing the $\max$ function to switch between them. For this reason, the default solver remains the bracketed binary mode-count search, and the Brent path is treated as an exploratory option with binary fallback rather than as the evidentiary backbone of the paper.

\textbf{Component decomposition.} The \texttt{detect\_components()} function splits the data at the KDE trough and computes per-side statistics. This is a fast and interpretable heuristic, but it lacks the statistical rigor of parametric approaches such as Gaussian mixture models (GMM) fitted via expectation-maximization with BIC selection (\S\ref{sec:gmm}). The trough-split method can be biased when the true components overlap substantially or have unequal variances, and it does not provide uncertainty estimates for the component parameters. We recommend it as an exploratory tool, with GMM as the preferred option for downstream parameter estimation in hard cases.

\textbf{Bimodality strength thresholds.} The \texttt{bimodality\_strength()} metric $h_{\text{crit}} / \hat{h}_{\text{silverman}}$ provides a useful continuous measure, but the weak/moderate/strong labels are heuristic summaries of the current benchmark set rather than calibrated decision rules. They are useful for descriptive reporting, but they should not be read as formal cutoffs.

\textbf{Bootstrap confidence intervals.} The bootstrap intervals for $h_{\text{crit}}$ (\S\ref{sec:bootstrap_ci}) are useful for exploratory uncertainty support, but they should still be treated cautiously until broader coverage studies are available. The current implementation uses a manual percentile bootstrap with failure accounting and explicit provenance; that is an improvement over the earlier percentile-only version, but it does not by itself justify strong inferential language.

\textbf{Large-sample testing.} Exact nonparametric tests can become expensive for datasets with $n > 5000$. For large-sample analyses, users may prefer Silverman's bootstrap test or a smaller \texttt{n\_boot} setting, and the implementation now warns when the default bootstrap count is used on large samples.

\textbf{Additional limitations.} The current implementation defaults to the Gaussian kernel, the standard choice in much of the multimodality literature, while also exposing additional kernel options for robustness checks. The bootstrap implementation does not yet support parallel execution across multiple cores, a natural extension for large-scale simulation studies. Finally, an adaptive bandwidth selection method that performs well on unequal-weight mixtures would be a valuable addition, potentially combining Silverman's rule with a pilot estimate of component structure.

\subsection*{Acknowledgments}

We gratefully acknowledge the developers of the \textbf{multimode} R package, whose work provided the validation benchmark for this project. We thank the open-source Python community for the foundational libraries (NumPy, SciPy) upon which \textbf{\method} is built.

\bibliography{refs}

\end{document}